\documentclass[12pt,preprint]{aastex}
\shorttitle{a mechanism for enhancing activity in
binaries}
\shortauthors{T. Zaqarashvili, G. Javakhishvili \& G.
Belvedere}
\begin{document}

\title{On a mechanism for enhancing magnetic activity in tidally
interacting binaries}
\author{T. Zaqarashvili}
\affil{School of Mathematics and Statistics,
University of St
Andrews, St Andrews, Fife KY16 9SS, Scotland\\
 Abastumani Astrophysical Observatory, Al.
Kazbegi ave. 2a,
380060 Tbilisi, Georgia}
\author{G. Javakhishvili}
\affil{Abastumani Astrophysical Observatory, Al.
Kazbegi ave. 2a,
380060 Tbilisi, Georgia}
\and
\author{G. Belvedere}
\affil{Dipartimento di Fisica e Astronomia, Universit\'a di
Catania, Via S.Sofia
78, I-95123 Catania, Italy}
\begin{abstract}
We suggest a mechanism for enhancing magnetic activity in
tidally interacting binaries. We suppose that the deviation of the primary star from
spherical symmetry due to the tidal influence of the
companion leads to stellar pulsation in its
fundamental mode. It is shown that stellar radial pulsation
amplifies torsional Alfv{\'e}n waves in a dipole-like magnetic field, buried in the interior, according to the recently proposed swing
wave-wave interaction \citep{zaq1}. Then amplified Alfv{\'e}n waves lead to the onset of large-scale torsional oscillations, and magnetic flux tubes arising
towards the surface owing to magnetic buoyancy
diffuse into the atmosphere producing enhanced chromospheric and coronal
emission.
\end{abstract}

\section{Introduction}

It is widely accepted that most
single stars and components of wide binary systems follow well-defined
rotation-activity relationships. However some stars - particularly components of
close binaries - show larger chromospheric and coronal activity in comparison
to similar single stars with the same rotation period. \citet{you77}
have noticed that binary giants with circular orbits have a tendency to show
enhanced CaII H and K emission, this stressing the importance of binarity in the
phenomenon of enhanced chromospheric emission. \citet{bas85} and \citet{sim87} noticed that subgiants in synchronized binary systems were more
active than single stars with the same rotation period. \citet{rut87}
attributes this "overactivity" to a difference in the internal stellar
structure and points out that overactive stars do not appear to deviate
from the flux-flux relationships, thus suggesting that their atmospheric
structure does not differ too much from that of other cool stars. \citet{gle88} noticed that MgII h and k fluxes in close binary systems correlate much
better with parameters connected to the separation of the components than with rotational parameters.

\citet{sch91} showed that the
chromospheric, transition-region, and coronal emissions from relatively close
binaries are enhanced as compared to single dwarfs and giants with the same
rotation period. They concluded that a star can be called overactive if the
radiative losses from its outer atmosphere are significantly over the
level expected from a single cool star with the same mass, chemical
composition, age and average surface rotational rate, while these radiative losses
do not deviate significantly from the flux-flux relationships defined for the class
of single cool stars. Overactivity appears to occur whenever a binary system
containing a cool primary or two cool components is sufficiently close to
induce strong tidal interaction.

\citet{med95} also showed that tidal effects play a direct role
in determining the X-ray activity level in binary evolved stars. The
circularisation of the orbit was a necessary property for enhanced coronal
activity. \citet{dem93} concluded that synchronous binaries show a
slight trend for increasing chromospheric emission with decreasing period,
while
the asynchronous binaries show abnormally high activity levels for their
rotational periods. Recently \citet{gun98} also point out the
influence of unknown effects of binarity on the activity levels.

Thus the observations stress the influence of the companion's gravity on
the magnetic activity of stars in relatively close binary systems.
In order to explain the phenomenon the mechanism of magnetic activity must be clearly understood.
It is widely believed that magnetic activity in the Sun and late
type solar-like stars can be explained in the framework of a mean-field dynamo
operating in or just below the convective zone \citep{par55a,par55b,ste66,yos78a,yos78b,bel91}.
However alternative theories, mainly some kind of hydromagnetic oscillator (or torsional oscillations) have been developed from time to time \citep{wal49,cow53,pid71,lay79,dic82,gou88}.
Although the mean-field dynamo seems to have captured the essentials of solar activity even if many uncertainties still remain (see, e.g., reviews by Belvedere 1985, 1997; Schmitt 1994, Brandenburg 1994),
the enhanced magnetic activity
in binaries can be hardly explained by a mean field dynamo, also because
dynamo requires large latitudinal or radial angular velocity gradients. In
fact observations show small latitudinal gradients, and almost nothing is known
about radial ones.
\citet{sch91} suggested a mechanism for the enhancement of dynamo efficiency related to the motion of stars about the system's centre of gravity, which lies well outside the stars. However no clear physical or mathematical formulations of the influence were developed after this suggestion and the problem remained unsolved.  Therefore it is natural to seek the key of the phenomenon in a possibly alternative theory for strong magnetic activity. The main problem of oscillator theories was the absence of an energy source to supply oscillations. If binarity can support torsional oscillations inside the star, then this may lead to a straightforward explanation of enhanced magnetic activity.

In this paper we develop the idea that stellar pulsation induced by the gravitational force of the companion star in tidally interacting binaries may amplify the torsional oscillations of a seed magnetic field, thus leading to enhanced chrmospheric emission. We suggest that the deviation of the star from spherical symmetry due to the gravitational influence of the companion induces stellar pulsation in its fundamental frequency, as it tries to retain its
original spherical form. Then we show that
the radial pulsation of the star with a dipole-like magnetic field causes
amplification of torsional Alfv{\'e}n waves in the interior. The mechanism of
coupling between radial pulsation and torsional Alfv{\'e}n waves is based on the
recently proposed swing wave-wave interaction \citet{zaq1}, which
accounts for the coupling between longitudinal and transversal waves. The
main physical meaning of the swing interaction is that compressible waves
cause periodical variation of both medium density and magnetic field (thus, of the Alfv{\'e}n speed),
affecting the propagation properties of the Alfv{\'e}n waves. Then the temporal evolution of pure Alfv{\'e}n waves is governed by Mathieu's equation and consequently the harmonics with half the frequency of compressible
waves grow exponentially in time. That means that the energy of
longitudinal waves can be transferred to transversal waves. The coupling between sound and Alfv{\'e}n
waves propagating along the magnetic field \citep{zaq1,zaq2} as well as
between fast magnetosonic waves propagating across the magnetic field and
Alfv{\'e}n waves propagating along the field \citep{zaq02}, is
studied in the simplest rectangular case. It is also supposed that stellar
radial pulsation can amplify torsional Alfv{\'e}n waves, which, in certain
conditions, may lead to the onset of torsional oscillations of a
dipole-like magnetic field \citep{zaq01}. In the present paper we show in details how the
pulsation induced by the tidal distortion of the star can amplify the torsional
Alfv{\'e}n waves.

\section{Resonant torsional Alfv{\'e}n waves in the
stellar interior}
Consider a binary system which is well separated so
that
mass transfer by Roche-lobe overflow does not occur.
Then we can exclude mass transfer as a mechanism for enhanced chromospheric
activity. Let the primary be
the primary as a main sequence star (or a subgiant) with a
convective envelope. The secondary can be a different type of star (if it is also a
main sequence or subgiant star, then the same mechanism can be
applied to both
components of the binary system). We will concentrate on
the dynamics
of the primary considering the gravity of the companion as an
external
force.

We use the ideal magnetohydrodynamic (MHD) equations:
\begin{equation}
{{{\partial {\rho}}}\over {\partial t}} +
{\nabla}({\rho}{\bf v})=0,
\end{equation}\begin{equation}
{\rho}{{{\partial \bf v}}\over {\partial t}} +
{\rho}({\bf v}{\cdot}{\nabla})
{\bf v} = - {\nabla}p + {\bf j}{\times}{\bf B} -
{\rho}{\nabla}{\phi},
\end{equation}\begin{equation}{\nabla}{\times}{\bf
B}={\mu}{\bf j},
\end{equation}
\begin{equation}
{{{\partial \bf B}}\over {\partial
t}}={\nabla}{\times}({\bf v}{\times}{\bf B}),
\end{equation}
\begin{equation}{\nabla}^2{\phi}=4{\pi}G{\rho},
\end{equation}
\begin{equation}
p=p_0\left ({{\rho}\over {\rho_0}}\right )^{\gamma},
\end{equation}
where $\rho$ is the medium density, $p$ is the
pressure, ${\bf v}$
is the velocity, $\bf B$ is the magnetic field, $\bf
j$ is the
current, G is the gravitational constant, $\phi$ is the gravitational potential, $\mu$ is
the magnetic
permeability and $\gamma$ is the ratio of specific heats. We consider a medium with zero viscosity, infinite
conductivity and negligible displacement current.
We argue that the coupling between pulsation and torsional
waves occurs below the convection zone, therefore we do not consider convective effects and consequently the $\alpha$-term in the induction equation. Here and in remaining part of the paper we also
neglect the rotational effects and consider a spherically
symmetric density distribution.
The gravitational influence exerted by the companion is also neglected in the equations, because we are interested to the companion's gravity only as a source for the initial deviation from the equilibrium. Its action during the further processes studied here is not relevant, so it is  neglected for simplicity.

We consider an unperturbed dipole-like seed magnetic field in the stellar interior. The existence of such stable large-scale configuration in the interior is an open question. A wide class of magnetic field topologies that are either purely toroidal or purely poloidal have been shown to be dynamically unstable \citep{tay73,wri73,mar73}. However special stable configurations with poloidal and toroidal field of similar strength might exist \citep{tay80,mes84}. The magnetic fields of magnetic A stars and the magnetic white dwarfs, which do not change on long time scales, are the observational evidence of stable configurations \citep{spr99}. In general it may be stated that curl-free magnetic configurations are stable while electrical currents leads to instability. All magnetic configurations studied by Tayler are current carrying, so they automatically lead to instability due to the Lorentz force. However if the magnetic field inside a star originates from the interstellar field during the gravitational contraction, then the Lorentz force acting on the induced currents during the contraction may redistribute the magnetic field leading to a minimum potential energy state (force-free configurations) which is relatively stable (see also \citet{tay86} for a similar phenomenon).
But the process of large-scale magnetic field formation is beyond the scope of this paper and therefore for simplicity we consider a purely poloidal magnetic field which in spherical coordinates ($r,\theta,\phi$) in the equatorial regions may be written as
\begin{equation}
{\bf B} = (0,B_{\theta}(r),0).
\end{equation}
It must be mentioned that all the results obtained in the paper are similar also for a purely toroidal unperturbed magnetic field, therefore the physical meaning of the phenomenon does not depend on the magnetic configuration.

We also consider a relatively weak magnetic field, so that ${\bf M}/{\mid{\bf
\Upsilon}\mid}{\ll}1$, where $\bf M$ is the magnetic and $\bf \Upsilon$ is gravitational potential energy. Then the star can be considered to be in hydrostatic equilibrium (if we neglect the companion's gravity) so that its own gravity is balanced by the pressure gradient.

The slight deviation from the equilibrium leads to stellar pulsation which may be studied by the linear perturbation theory. The stellar
adiabatic oscillations (without the magnetic field) can be divided into two classes \citep{cox80,gau95}: high-frequency p-modes or pressure modes, the restoring force of which is the pressure gradient, and low-frequency g-modes, where  buoyancy acts as the restoring force. There are also intermediate oscillations sometimes called f-modes \citep{cow41}.

Any external action on the bounded system leads to oscillation in the first eigenfrequency which is also called the fundamental frequency of the system. The mechanical analog of the fundamental oscillation is the tuning fork in the case of an impulsive force and the musical trumpet in the case of a continuous force. Additionally a periodic external force leads to resonance when its frequency equals one of the system eigenfrequencies.

The oscillation of the star in the fundamental mode yields a wavelength comparable with the stellar radius. Therefore the fundamental frequency will be of order of ${\sqrt {GM/R^3}}$, where  $R$ and $M$ are the radius and the mass of star  \citep{cow52} (for the Sun the oscillation period is about 2 hours).
A purely radial pressure mode ($l=0$, where $l$ is the spherical degree of a mode) with one velocity node at the centre, and one antinode at the surface will have a frequency of the order of $c_0/R$ (${\sqrt {GM/R^3}}{\sim}c_0/R$ because of the equilibrium between pressure gradient and gravity), where $c_0$ is the mean sound speed in the interior.
A large-scale magnetic field inside the star probably causes the frequency splitting  corresponding to fast and slow MHD waves \citep{rob83}. If the magnetic field is relatively weak then the frequency corresponding to fast waves remains almost the same $~c_0/R$, while the frequency corresponding to slow MHD waves will be of order of $v_A/R$, where $v_A$ is the mean Alfv{\'e}n speed in the interior.

\subsection{Oscillation of the primary induced by
tidal distortion}

The tidal force exerted by the companion leads to the
deviation of the star from equilibrium. In synchronous binaries
the deformation is always in the direction of the companion, because of the same value of the rotational and orbital period. Then for circular orbits a new equilibrium may be established where the tidal force plays a significant role. This is a stationary tide. However in asynchronous binaries the tidal bulk will change location permanently and a stationary state can be hardly established even for circular orbits. This is a dynamical tide. Now the tidal bulk will be "retarded" with respect to the companion. This retarded tide may lead to stellar oscillation when the star tries to return to its spherically symmetric form (due to this retarded tide exerted from the Moon, the Earth's rotation period is supposed to be increasing). Like the musical trumpet (or the tuning fork) which oscillates in its fundamental
frequency under the action of an external force, the primary
probably will oscillate in its fundamental mode. The stronger is the tidal action of the companion, the stronger is the oscillation.

If the reciprocal of the orbital period is close to the eigenfrequencies of the free oscillations of the star, then resonance can lead to significant enhancement of the oscillation \citep{cow41,zah70,sme98}. But the resonant conditions can be fulfilled only for low-frequency g-modes. Moreover, the observational evidence of tidally excited resonant oscillations is not well established, possibly due to the absence of any systematic study \citep{wil02}.
However, a retarded tidal bulk in asynchronous binaries may lead to the oscillation of the primary star in the fundamental frequency which is higher than the orbital angular frequency.

In any case, it may be supposed that the star under consideration in tidally interacting binaries oscillates either in its fundamental frequency or in a low frequency g-mode which is in resonance with the tidal force.
For simplicity
(and avoiding the uncertainty connected with the velocity polarization of the waves) we consider a coupling between purely radial axisymmetric pulsation and
purely torsional Alfv{\'e}n waves, thus we take ${\partial/{\partial {\phi}}}=0$. Of course, realistic conditions are much more complicated (tidally induced oscillations probably depends on $\phi$), however for a better understanding of the coupling mechanism it is worth looking at the simplest case.

Then the linearized equations (1)-(6) can be split into radial and toroidal components, where the radial component corresponds to pulsations and the toroidal component to torsional waves.

The radial part of the equations is
\begin{equation} {{{\partial b_{\theta}}}\over
{\partial t}}=
-B_{\theta}{{{\partial u_r}}\over {\partial r}} -
{{{\partial
B_{\theta}}}\over {\partial r}}u_r -
{{B_{\theta}}\over r}u_r,
\end{equation}\begin{equation}
{{{\partial {\delta}{\rho}}}\over {\partial t}}=
-\rho_0{{{\partial
u_r}}\over {\partial r}} -
{{{\partial {\rho_0}}}\over {\partial r}}u_r -
2{{\rho_0}\over r}u_r,
\end{equation}\begin{equation}
{\rho_0}{{{\partial u_r}}\over {\partial t}} = -
{{{\partial}}\over
{\partial r}}\left[{\delta}p +
{{B_{\theta}b_{\theta}}\over
{{\mu}}}\right ] -{{2B_{\theta}b_{\theta}}\over
{{\mu}r}} - {\delta
{\rho}}{{{\partial {{\phi}}}}\over {\partial
r}},\end{equation}
\begin{equation}{{{\partial {\delta}p}}\over {\partial
t}} + u_r{{{\partial
{p_0}}}\over {\partial r}}= c_0^2\left ({{{\partial
{\delta}{\rho}}}\over
{\partial t}} + u_r{{{\partial {{\rho}_0}}}\over
{\partial r}} \right ),
\end{equation}
where $c_0=\sqrt{{\gamma}p_0/{\rho}_0}$ is the sound
speed, while the
toroidal part is\begin{equation}{{{\partial
}{b_{\phi}}}\over {{\partial}t}}=
{{B_{\theta}}\over r}{{\partial u_{\phi}}\over
{\partial {\theta}}},
\end{equation}\begin{equation}
{\rho_0}{{{\partial u_{\phi}}}\over {\partial t}} =
{{B_{\theta}}\over
{{\mu}r}}{{\partial b_{\phi}}\over {\partial
{\theta}}}.\end{equation}
$u_r, u_{\phi}, b_{\theta}, b_{\phi}, {\delta}{\rho}$
and ${\delta}p$ are the
velocity, magnetic field, density and
pressure pertubations respectively.
Equations (8)-(11) govern the radial pulsation of
the star, while
equations (12)-(13) describe the torsional Alfv{\'e}n
waves. It must be noticed that if an unperturbed purely poloidal magnetic field (7)
is replaced by a purely toroidal field ${\bf B}=(0,0,B_{\phi}(r))$, equations (8)-(13) remain unchanged. The Alfv{\'e}n waves now will be described by $b_{\theta},\,u_{\theta}$ instead of $b_{\phi},\,u_{\phi}$. Therefore the phenomenon does not depend on the particular magnetic field configuration.

In general, stellar radial adiabatic pulsation can be represented as
a standing spherical wave \citep{cox80, bir82, rob83, chr88}, with a linear radial velocity field
\begin{equation}u_r={\alpha}F(r){\sin}(\omega_nt),
\end{equation}
where $\omega_n$ is the eigenfrequency, $F(r)$ is the
eigenfunction and $\alpha$ is the pulsation amplitude.
The expression of the eigenfunction $F(r)$ depends on the spatial profiles of the unperturbed values and may be represented by some combination of spherical Bessel functions. However, here we are not interested in
specific pulsation functional forms, but our aim is to show that pulsation in general leads to amplification of torsional Alfv{\'e}n waves. Therefore we retain its general form $F(r)$, which then may be specified by choosing the spatial distribution of pressure, density and magnetic field throughout the star. However it must be mentioned that equations
(8)-(11) govern only the fast mode of pulsation, because the velocity is strictly across the field lines.
The equations governing the slow mode of pulsation
must also concern the radial component of the magnetic field $B_r$, which significantly complicates the problem. Therefore we describe only the coupling between the fast
mode of pulsation and the torsional Alfv{\'e}n waves, keeping in mind that a similar physical process
occurs for the slow mode too.

Using expression (14) we find the density
and magnetic field perturbations from the linearized
continuity and induction equations
(8)-(9):
\begin{equation}
\delta\rho={{\alpha}\over
{\omega_n}}{\cos}(\omega_nt)F_{\rho}(r),\end{equation}
\begin{equation}
{b_{\theta}}={{\alpha}\over
{\omega_n}}{\cos}(\omega_nt)F_{b}(r),\end{equation}
where\begin{equation}
F_{\rho}(r)={\rho_0(r)}{{\partial F(r)}\over
{\partial r}} + \left ({{\partial {\rho_0(r)}}\over
{\partial r}}
+ {{2{\rho_0(r)}}\over r}\right
)F(r),\end{equation}\begin{equation} F_{b}(r)=
B_{\theta}(r){{\partial F(r)}\over {\partial r}} +
\left
({{\partial {B_{\theta}(r)}}\over {\partial r}} +
{{{B_{\theta}(r)}}\over r}\right )F(r).\end{equation}
Equations (15)-(16) show that the radial pulsation
leads to local
periodical variation of density and magnetic field
throughout the star. It must be mentioned that the fundamental mode of pulsation can be interpreted as a standing wave with one velocity node at the stellar centre and one antinode at the surface, which imposes the corresponding boundary conditions on the eigenfunction $F(r)$.

The variation of magnetic field and density due to pulsation leads to  periodical variation of the Alfv{\'e}n speed at each level $r$. In next section we show that it determines exponential amplification of torsional Alfv{\'e}n waves with half the frequency of pulsation.

\subsection{Swing amplification of torsional Alfv{\'e}n
waves due to the
radial pulsation of the star}

The good mechanical analogue of swing coupling between longitudinal and transversal waves is a
mathematical pendulum with a stiffness spring \citep{zaq02}.
The system includes two kinds of oscillations: the pendulum transversal oscillations due to gravity and the
spring oscillation along the pendulum axis due to stiffness force. The
oscillations are coupled and it is shown that in certain conditions, when the
frequency of spring oscillations is twice as large as that of pendulum
transversal oscillations, the energy of spring oscillations along the pendulum
axis is transferred to transversal oscillations
and {\it vice versa}.

In the case of a star, the pulsation corresponds to the spring oscillation and
the torsional Alfv{\'e}n waves correspond to the transversal oscillations. The
torsional Alfv{\'e}n waves are the result of the magnetic tension force action
against the fluid inertia. On the other hand, the pulsation makes the work against the tension force during each compression and thus may lead to  amplification of the transversal oscillations with half the frequency of the pulsation \citep{zaq01,zaq02}.

In order to study the influence of pulsations on Alfv{\'e}n waves
we must retain the
expressions of $u_r$, $\delta\rho$ and $b_{\theta}$ in
equations
(12)-(13) which now take the form
\begin{equation}
{{{\partial }{b_{\phi}}}\over {{\partial}t}}=
{{B_{\theta} + b_{\theta}}\over r}{{\partial
u_{\phi}}\over {\partial
{\theta}}} - {{\partial u_r}\over {\partial
r}}b_{\phi}- {{u_r}\over r}b_{\phi},
\end{equation}\begin{equation}
({\rho_0} + {\delta \rho}){{{\partial u_{\phi}}}\over
{\partial t}} +
({\rho_0} + {\delta \rho}){{u_r}\over r}u_{\phi}=
{{B_{\theta} +
b_{\theta}}\over {{\mu}r}}{{\partial b_{\phi}}\over
{\partial {\theta}}}.
\end{equation}
From these equations we derive the Hill type second
order differential
equation with periodical coefficients
\begin{displaymath}
{{\partial^2 u_{\phi}}\over {\partial t^2}} + \left
({{\partial
u_r}\over {\partial r}} + {{2u_r}\over r} +
{1\over {\rho_0 + \delta\rho}}{{\partial \delta{\rho}}\over {\partial t}} -
{1\over {B_{\theta} + b_{\theta}}}{{\partial b_{\theta}}\over {\partial t}}\right )
{{\partial u_{\phi}}\over {\partial t}} -
\end{displaymath}
\begin{equation}
- \left [{{(B_{\theta} + b_{\theta})^2}\over {\mu({\rho_0}
-\delta{\rho})}r^2}{{{\partial^2}\over {\partial
{\theta}^2}}}
+ {1\over r}{{\partial u_r}\over {\partial t}} + {1\over
{B_{\theta} + b_{\theta}}}{{u_r}\over
r}{{\partial b_{\theta}}\over {\partial t}} - {1\over {\rho_0 +
\delta\rho}}{{u_r}\over
r}{{\partial \delta{\rho}}\over {\partial t}} - {{u_r}\over r}{{\partial u_r}\over
{\partial r}} -
{{u_r^2}\over r^2}\right ]u_{\phi}=0.
\end{equation}
Using the well known transformation
\begin{equation}
u_{\phi} = u(t)\exp \left ({-{1\over 2}{\int \left [{{\partial
u_r}\over {\partial r}} + {{2u_r}\over r} +
{1\over {\rho_0 + \delta\rho}}{{\partial \delta{\rho}}\over {\partial t}}
-{1\over {B_{\theta} + b_{\theta}}}{{\partial b_{\theta}}\over {\partial t}}\right ]dt}}\right )
\end{equation}
leads to elimination of the first time derivative and after neglecting terms of order $\alpha^2$ we have
\begin{displaymath}
{{\partial^2 u}\over {\partial t^2}} -{{V_A^2(r)}\over
r^2}\left [1+
{{\alpha}\over {\omega_n}}
\left (2{{F_b(r)}\over {B_{\theta}(r)}} -
{{F_{\rho}(r)}\over
{\rho_0(r)}}\right ){\cos}(\omega_nt)\right
]{{\partial^2 u}\over {\partial
{\theta}^2}} +
\end{displaymath}
\begin{equation}
 + {{{\alpha}{\omega_n}}\over 2}\left [-{{\partial
F(r)}\over {\partial r}} + {{F_{\rho}(r)}\over
{\rho_0(r)}} - {{F_{b}(r)}\over {B_{\theta}(r)}}\right ]{\cos}(\omega_nt)u=0,
\end{equation}
where
\begin{equation}
V_A(r)={{B_{\theta}(r)}\over {{\sqrt
{\mu{\rho_0(r)}}}}}
\end{equation}
is the Alfv{\'e}n speed.
Equation (23) has time-dependent coefficients
while the radial
coordinate $r$ stands as a parameter. Then we can
perform a Fourier transform with $\theta$
dependence
\begin{equation}
u=\int{{\hat u(l,t)}e^{il\theta}dl},
\end{equation}
which leads to Mathieu's equation
\begin{equation}
{{\partial^2 {\hat u}}\over {\partial t^2}} + \left
[V_A^2(r){{l^2}\over r^2} +
{\alpha}{\Psi}(r){\cos}(\omega_nt)\right ]{\hat u}=0,
\end{equation}where
\begin{equation}
{\Psi}(r) = {{V_A^2(r)l^2}\over
{{\omega_n}r^2}}\left
(2{{F_{b}(r)}\over {B_{\theta}(r)}} -
{{F_{\rho}(r)}\over {\rho_0(r)}} \right )
- {{{\omega_n}}\over 2}\left ({{\partial F(r)}\over
{\partial r}} - {{F_{\rho}(r)}\over {\rho_0(r)}} +
{{F_{b}(r)}\over {B_{\theta}(r)}}\right ).
\end{equation}
This equation is well studied and its main resonant
solution occurs when
the frequency of the Alfv{\'e}n waves is half that of the
external force
\begin{equation}{\omega_A}={{B_{\theta}(r)l}\over
{r{\sqrt{\mu{\rho_0(r)}}}}}={{\omega_n}\over
2}.\end{equation}
In this case equation (26) has the exponentially growing solution
\citep{lan88}

\begin{equation}
{\hat u}(t)={\hat u}_0e^{{{\left
|{\alpha}{\Psi(r)}\right |}\over
{2\omega_n}}t}\left
[{\cos}{{\omega_n}\over 2}t {\mp}
{\sin}{{\omega_n}\over 2}t \right ],
\end{equation}
where ${\hat u}_0={\hat u}(0)$ and the
phase sign
depends on ${\alpha}{\Psi(r)}$; it is $+$ for negative
${\alpha}{\Psi(r)}$ and $-$ for positive
${\alpha}{\Psi(r)}$. Note that the solution has a resonant character within the frequency interval

\begin{equation} {\left |{\omega_A} -
{{\omega_n}\over 2}
\right |}<{\left |{{\alpha\Psi}\over {\omega_n}}
\right |}.
\end{equation}

Thus pulsation leads to exponential amplification
of torsional Alfv{\'e}n waves at half frequency
${1\over 2}{\omega_n}$.
The growth rate depends on the amplitude $\alpha$ and spatial
structure of the pulsation eigenfunction $F(r)$, which in turn depends on the radial structure of density $\rho_0$ and
magnetic field $B_{\theta}$.
The resonant condition (28) imposes a restriction on the
wave number $l$ at each distance
$r$ from the stellar center. In other words, the pulsation "picks up"
the harmonics of torsional Alfv{\'e}n waves with a certain
wavelength at each
distance $r$. It gives the idea that at some distances from the
stellar center, where the conditions for the onset of standing waves are satisfied
\begin{equation}
l={{\omega_n}\over
2}{{r{\sqrt{\mu{\rho_0(r)}}}}\over
{B_{\theta}(r)}}=1,2,3...,
\end{equation}
the amplified Alfv{\'e}n waves may lead to the setup of
torsional oscillations. In these regions the pulsation energy can be
dramatically "absorbed" by transversal oscillations.

It is clear that the energy transferred into torsional Alfv{\'e}n waves is to be extracted from pulsation. In fact the term $-{b^2_{\phi}/{\mu}r} - {{\partial {(b^2_{\phi}/{2\mu}})}/{\partial r}}$ (or ponderomotive force), which describes the back reaction of Alfv{\'e}n waves on pulsation, should be added in the righthand-side of the equation (10). At the initial stage,
this term is of second order strength, however it becomes significant after some time because of the exponential growth of $b_{\phi}$ and leads to the damping of pulsation. This process is clearly seen in the rectangular case \citep{zaq02}.

\section{Discussion}

Observations show enhanced chromospheric and coronal
activity in relatively close binaries \citep{sch91}.
The binary systems considered in that paper are well separated so that  mass transfer by
Roche-lobe overflow does not occur. The chromospheric, transition-region and
coronal emissions from the binaries are enhanced in comparison to single
stars with the same mass, chemical composition, age and mean surface rotation
rate. This is somehow a strange phenomenon, because in the framework of dynamo theory the magnetic activity does not
depend on whether the star is single or component of a binary system.

It is clear however that tidal interaction somehow causes the enhancement of magnetic activity. It leads to the idea that
some mechanism, other than classical dynamo, may generate in the stellar
interior the magnetic fields that give rise to the strong activity
observed in the considered binary systems.Several alternative mechanisms were supposed to explain magnetic activity in the Sun and late type stars (see the review of \citet{bel85}). Most of them are connected to the hydromagnetic oscillation of a seed magnetic field either in the core  \citep{dic82,gou88} or in the radiative layers \citep{wal49,cow53,pid71,lay79,zaq97}. However, a drawback of these mechanisms is the absence of an energy source to support the oscillations. The recently suggested mechanism \citep{zaq01} of transformation of pulsation energy into energy of torsional oscillations is the first step towards this direction. If this is the case, then any energy source able to support the pulsation can be also considered as the source of torsional oscillations.

Indeed, here we suggest that torsional oscillations, amplified by tidally induced pulsation, may be responsible for enhanced magnetic activity in tidally interacting binaries. The proposed mechanism is alternative to the mean field dynamo, therefore the
presence of a dynamo field is not considered. It is also clear that in our paper
we suggest a self-consistent mechanism which assumes a seed field in the radiative interior,
but operates differently from the alternative mechanisms suggested by the authors quoted above
which, on the other hand, refer only to the Sun. \\
We suppose that the deviation of the star from spherical symmetry
due to tidal interaction sets up its pulsation
in the fundamental mode like a tuning fork. The companion's gravity may induce stellar pulsation either due to resonance between a dynamic tide and g-mode oscillations
\citep{cow41,zah70,sme98, wil02} or to an unstable retarded tide in asynchronous systems.
The conditions required for resonance between the periodic tidal force and the stellar free oscillations can be fulfilled only for low-frequency g-modes. The fundamental frequency of pulsation is usually higher (for the Sun, it corresponds to a few hours) than the orbital angular frequency of binary systems. Although observational evidence of tidally excited resonant oscillations is not well established, possibly due to the absence of any systematic study \citep{wil02}, on the other hand, a retarded tide in asynchronous binaries will be unstable and may lead to stellar pulsation in the fundamental mode.

However the existence of asynchronous orbits in relatively close
binary systems is under question. The tidal theory \citep{zah77} predicts that spin-orbit synchronization of a component of a late-type close binary system occurs before the orbit becomes circular, unless the spin angular momentum is comparable to the orbital one \citep{hut80, sav84}. However, several
binary systems TZ For, $\lambda$ And, AY Cet and $\alpha$ Aur include
asynchronously rotating giants in circular orbits \citep{and84,hal86}. A very interesting RS CVn-type system is $\lambda$ And, which is
asynchronous and shows enhanced activity \citep{don95}. For instance, \citet{hab89} try to explain asynchronous rotation in close binary systems with circular orbits, however the problem still remains.

Thus we suggest that the tidal force exerted by the companion on the primary in binaries may induce stellar radial pulsation either due to resonance or due to a retarded tide. The next question is how pulsation may affect magnetic activity.

Suppose the star has large-scale dipole-like seed magnetic field in the interior. As we already discussed, the various magnetic configurations, mostly purely toroidal or purely poloidal, were found to be dynamically unstable
\citep{tay73,wri73,mar73}. However special stable configurations with poloidal and toroidal field of similar strength might exist \citep{tay80,mes84}.
Also plasma relaxation theory \citep{tay86} yields that instabilities may lead to a particular minimum-energy state (mainly force-free), which is relatively stable. Therefore for simplicity we consider the magnetic field to be purely poloidal. Notice that a purely toroidal magnetic field in the equatorial regions leads to the same equations, which indicates that the physics of the phenomenon does not depend on the particular field configuration.
The strength of the unperturbed magnetic field in the interior cannot be evaluated since the analysis is linear.
However, let us try to give an estimate of the energy which can be transferred
to magnetic oscillations due to tidal interaction, and of the mean magnetic field strength. Let the binary system components have mass $M$ and radius $R$ of the order
of the solar ones. The energy of tidal interaction is
${{GM^2R}/d^2}$, where $d$ is the orbital separation. Then, for an orbital separation $d{\sim}100R$, this energy is of the order of $10^{45}$ ergs.
The total magnetic field energy is expressed by: $B^2/2{\mu}{\cdot}(4/3){\pi}R^3$.
If we assume that all the tidal interaction energy goes into magnetic
field energy, we get a mean equipartition magnetic field
$B\sim10^6$ G, but this is clearly an upper limit. Indeed this result is obtained
in the ideal case that the whole tidal interaction energy is converted into the
magnetic one. For a more realistic situation let us suppose that, due
to viscous and ohmic dissipation, the conversion efficiency be only $1\%$. Even with this low efficiency,
the magnetic field strength can attain values as high as $>10^5$ G, which
are comparable with those commonly believed to exist at the base of the convection zone of the Sun and active stars.
These values are not very different from the surface fields measured in very active stars that
may reach $10^4$ G or more.

In the presence of a large-scale magnetic field, pulsation induces a periodical variation of the local Alfv{\'e}n speed at each distance from the stellar centre, which in turn leads to a crucial influence on the dynamics of torsional Alfv{\'e}n waves. We found that the time behaviour of torsional Alfv{\'e}n waves is governed by Mathieu's equation (26), therefore the harmonics with half the frequency of pulsation grow exponentially in time. The growth rate of torsional Alfv{\'e}n waves depends on the amplitude of pulsation (see also \citet{zaq01}).
This is a very important result, because pulsations can be easily excited by any
non-electromagnetic force. Then the torsional oscillations may have a number of energy sources, which can be of importance for the study of stellar activity in general. However the discussion of these situations is beyond the scope of this paper, therefore we only consider magnetic activity in binary stars. Gravitational energy in binary systems is much larger than the energy involved in magnetic activity. Unfortunately, it is not possible to evaluate the amount of energy transferred to torsional oscillations, because the theory is basically linear.
However, it is reasonable to consider the back-reaction of amplified Alfv{\'e}n waves on  pulsation itself. Energy conservation yields that the amount of the energy transferred into  Alfv{\'e}n waves must be extracted from pulsation itself. Indeed, this is the case, as can be easily shown in rectangular geometry \citep{zaq02}. However, if pulsation has a continuous energy source (in our case, the companion's gravity), then Alfv{\'e}n waves will grow until their
amplitudes exceed a certain value, after which magnetic buoyancy probably leads to eruption of magnetic tubes at the surface. There, they may release their energy, this leading to enhanced chromospheric and coronal activity.

In our model, we expect chromospheric and coronal activity to be stronger in asynchronous binaries. Some observations seem to indicate it \citep{dem93}, however future establishment of a better link between theory and observations is needed.

Another interesting point is that the unperturbed magnetic field in the
stellar interior causes the splitting of the fundamental frequency of adiabatic pulsation into a fast and a slow mode. Both modes can be interpreted as spherical standing waves with one node in the center and one antinode at the surface. The
fast mode corresponds to fast magnetosonic waves, and, for weak magnetic
field, their frequency is of the order of $c_0/R$. The slow mode corresponds to slow magnetosonic waves, and their frequency is of the order
of $v_A/R$. The difference between these
frequencies depends on the ratio of hydrodynamic and magnetic pressures
and can be very large for weak magnetic fields. So a star with an even
weak magnetic field should have two pulsation timescales. This means that
resonant Alfv{\'e}n waves must also have two timescales, which correspond to
fast and slow modes. In this paper we consider the coupling between the fast
mode of pulsation and torsional waves, because the description of the fast mode
is relatively easy, while the slow mode of pulsation requires more complicated mathematical
formulation. However, the physical meaning of the coupling is similar, therefore we may
argue about the coupling between the slow mode of pulsation and torsional waves
without giving detailed calculations (however it is essential to study the
details of their coupling in future). The particular characteristic of the
coupling between the slow mode and torsional Alfv{\'e}n waves is that the latter have spatial scales comparable to the stellar radius, because Alfv{\'e}n and slow magnetosonic waves have similar phase velocities in the case
of a weak magnetic field. Therefore they may lead to the onset of
large-scale torsional oscillations.

\section{Conclusion}

In conclusion we summarize the process which gives rise to enhanced chromospheric and coronal activity in tidally
interacting binaries in the three steps:

I - tidal interaction causes deviation of the star from
spherical symmetry, which leads to stellar pulsation in its fundamental frequency;

II - pulsation amplifies torsional Alfv{\'e}n waves through the mechanism of swing wave-wave interaction which leads to onset of torsional
oscillations in the radiative interior;

III - magnetic flux tubes
erupted at the surface by magnetic buoyancy
enhance chromospheric and coronal emission.

However, the proposed mechanism needs future study. The details of the mechanism of amplified pulsation due to tidal interaction should be specified. It is also essential to give
a detailed mathematical formulation of the coupling between the slow
mode of pulsation and torsional Alfv{\'e}n waves, which may lead to
large-scale torsional oscillations of a dipole-like magnetic field.

\begin{acknowledgements} The authors are grateful to B. Roberts and A. F. Lanza for useful
discussions. This work was supported by NATO Collaborative Linkage grant PST CLG.976557. TZ was also supported by the Department of Physics and Astronomy, Astrophysics Section of Catania University and by the Royal
Society/NATO Postdoctoral Fellowship programme. The authors are indebted to an anonymous referee for positive and valuable suggestions.

\end{acknowledgements}

\end{document}